\documentclass[twocolumn,a4paper,10pt]{revtex4-1}
\usepackage{CJKutf8,upgreek,fancyhdr}
\usepackage{graphicx}
\usepackage{booktabs}
\usepackage{verbatim}
\usepackage{dcolumn}
\usepackage{amssymb,bm,mathrsfs,bbm,amscd}
\usepackage[tbtags]{amsmath}
\usepackage{lastpage}

\begin{document}
\begin{CJK}{UTF8}{gbsn}

\title{Injection orbit matching for CSNS/RCS}
\author{Xiaohan Lu$^{1,2}$
 \quad Ming-Yang Huang$^{1,2}$
 \quad Sheng Wang$^{1,2}$
 }

\email{wangs@ihep.ac.cn}
\affiliation{%
$^1$ Dongguan Institute of Neutron Science, Dongguan 523808, China\\
$^2$ {\bf } Institute of High Energy Physics, Chinese Academy of Sciences, Beijing 100049, China\\
}

\begin{abstract}
The Rapid Cycling Synchrotron(RCS) of China Spallation Neutron Source(CSNS) employs painting injection to achieve uniform beam distribution and suppress space-charge effects. In the painting injection, the mismatch between the injection beam orbit and circulating orbit could make the non-uniformity of beam distribution and emittance growth, which may lead to beam loss. To match the injection beam orbit and the circulating beam orbit, it is necessary to identify the orbit of injection beam relative to the circulating orbit in the injection point. However, it's hard to measure this relative injection orbit directly. Theoretically, the relative injection beam orbit can be deduced by measuring the turn-by-turn beam position of a single turn injection beam. However, the intensity of single turn injected beam is too low to be measured with sufficient signal-noise ratio by using beam position monitor(BPM). In this paper, two effective methods based on multi-turn injection and turn-by-turn BPM are given to perform the match between the injection and the circulating beam orbit. The simulation results and application in the beam commissioning of CSNS/RCS show the validity of the methods.
\begin{description}
\item[PACS numbers]{29.27.Ac, 29.20.Lq}
\end{description}
\end{abstract}

\maketitle

\section{Introduction}
\label{sec:Introduction}
\begin{figure}
\includegraphics[width=8cm]{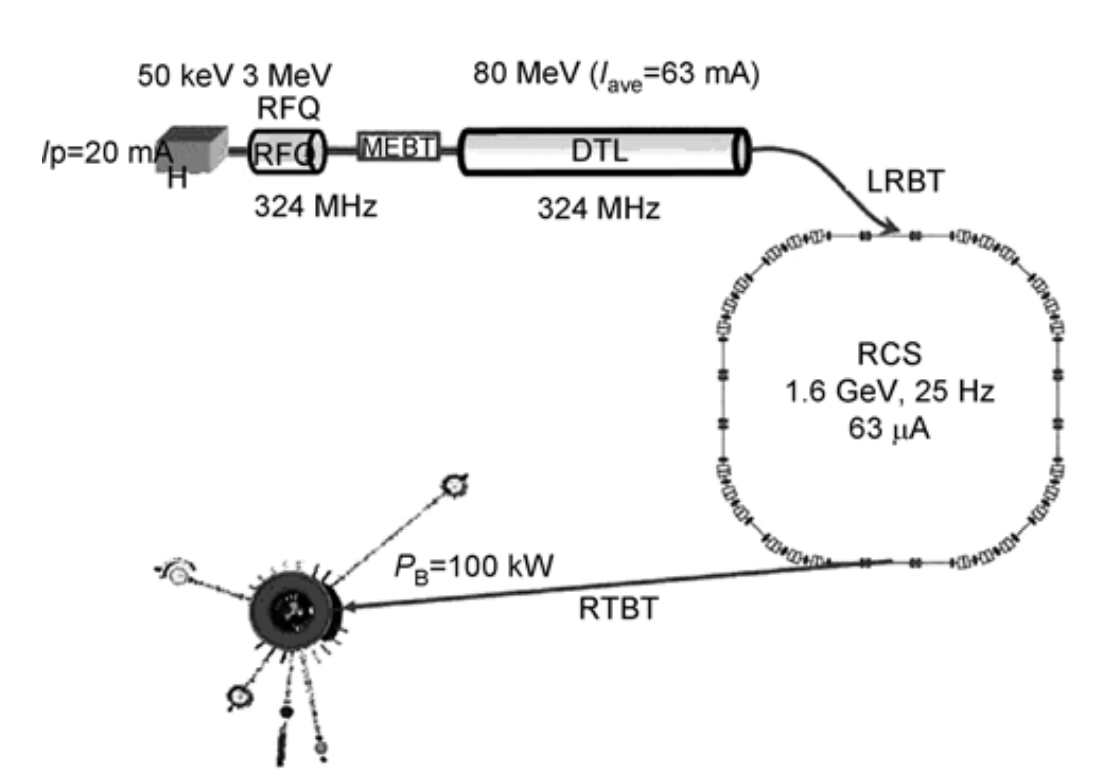}
\caption{\label{fig1} The schematic layout of CSNS}
\end{figure}

\begin{figure}
\includegraphics[width=8cm]{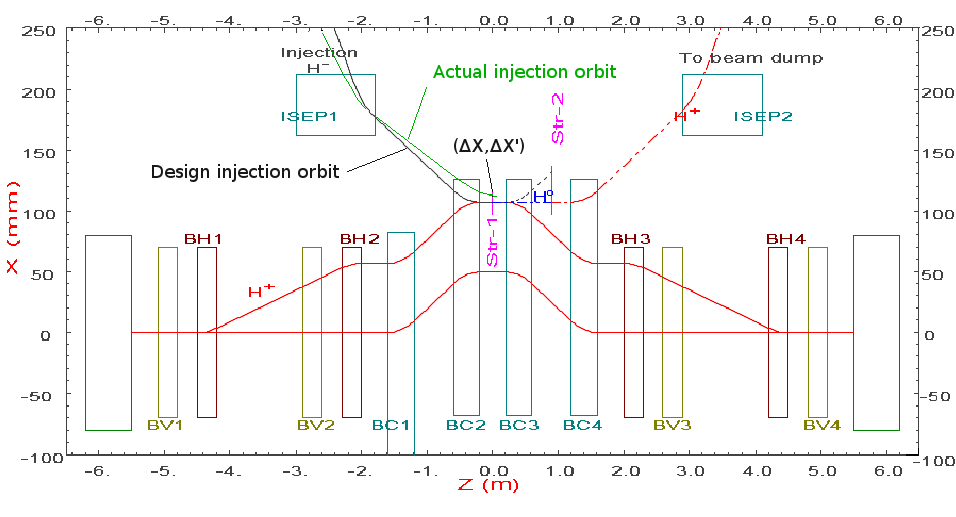}
\caption{\label{figinjlayout} (Color)The layout of injection system}
\end{figure}

China Spallation Neutron Source(CSNS) is a high intensity proton accelerator based 100 $kW$ pulsed neutron source~\citep{ref1}, its accelerators consist of an 80 $MeV$ H-linac and a 1.6 $GeV$ proton Rapid Cycling Synchrotron(RCS) with repetition rate of 25 $Hz$~\citep{ref2}. The schematic layout is shown in Fig.~\ref{fig1}, the main parameters are shown in Table~\ref{tab1}. The layout of injection system is shown in Fig.~\ref{figinjlayout}, where eight painting bump magnets (BH1$\sim$BH4 for horizontal and BV1$\sim$BV4 for vertical), and four chicane bump magnets (BC1$\sim$BC4) are accommodated in an uninterrupted straight section~\citep{ref3-1}~\citep{ref3}. The RCS employs painting injection to achieve uniform beam distribution and suppress the space-charge effects. 

To control the painting process precisely, it is necessary to identify the orbit of injection beam relative to the circulating orbit, which are $(\Delta x/\Delta y, \Delta x'/\Delta y')$ in transverse phase space, in the injection point. Hereafter we choose $(\Delta x, \Delta x')$ as the study object. With the identified relative injection orbit, the injection orbit and circulating orbit can be well matched upon the painting pattern. The mismatch between the injection beam orbit and circulating orbit could make the non-uniformity of beam distribution and emittance growth, which may lead to beam loss. Some simulations were done to show the effects of the injection orbit mismatch on the emittance growth and beam loss. In the simulation, taking the painting process of CSNS as an example, the anti-correlated painting process is simulated by using the particle tracking code ORBIT~\citep{ref4}. 

\begin{table}
\caption{ \label{tab1}  RCS Design Parameters.}
\begin{ruledtabular}
\begin{tabular*}{80mm}{l@{\extracolsep{\fill}}c}
\textbf{Parameters} & \textbf{Value} \\
\colrule
Circumference  & 227.92 m  \\
Superperiodicity & 4  \\
Injection energy & 80 Mev \\
Single injection beam peak current  & 15 mA \\
Extraction energy & 1.6 Gev \\
Repetition rate & 25 Hz \\
Extraction beam power & 100kW \\
Design Betatron tunes(h/v) & 4.86/4.78 \\
\end{tabular*}
\end{ruledtabular}
\end{table}

In the anti-correlated painting, horizontal painting fills RCS acceptance from center to outer phase space, and the matching conditions are $\Delta x=0, \Delta x'=0$. In the simulation, let $\Delta x'=0$, and sweep $\Delta x$ from $0~mm$ to $10~mm$. The turn number of the injection painting process was 200 and the first 2000 turns in acceleration process were considered in simulation.

\begin{figure}
\includegraphics[width=8cm,height=5cm]{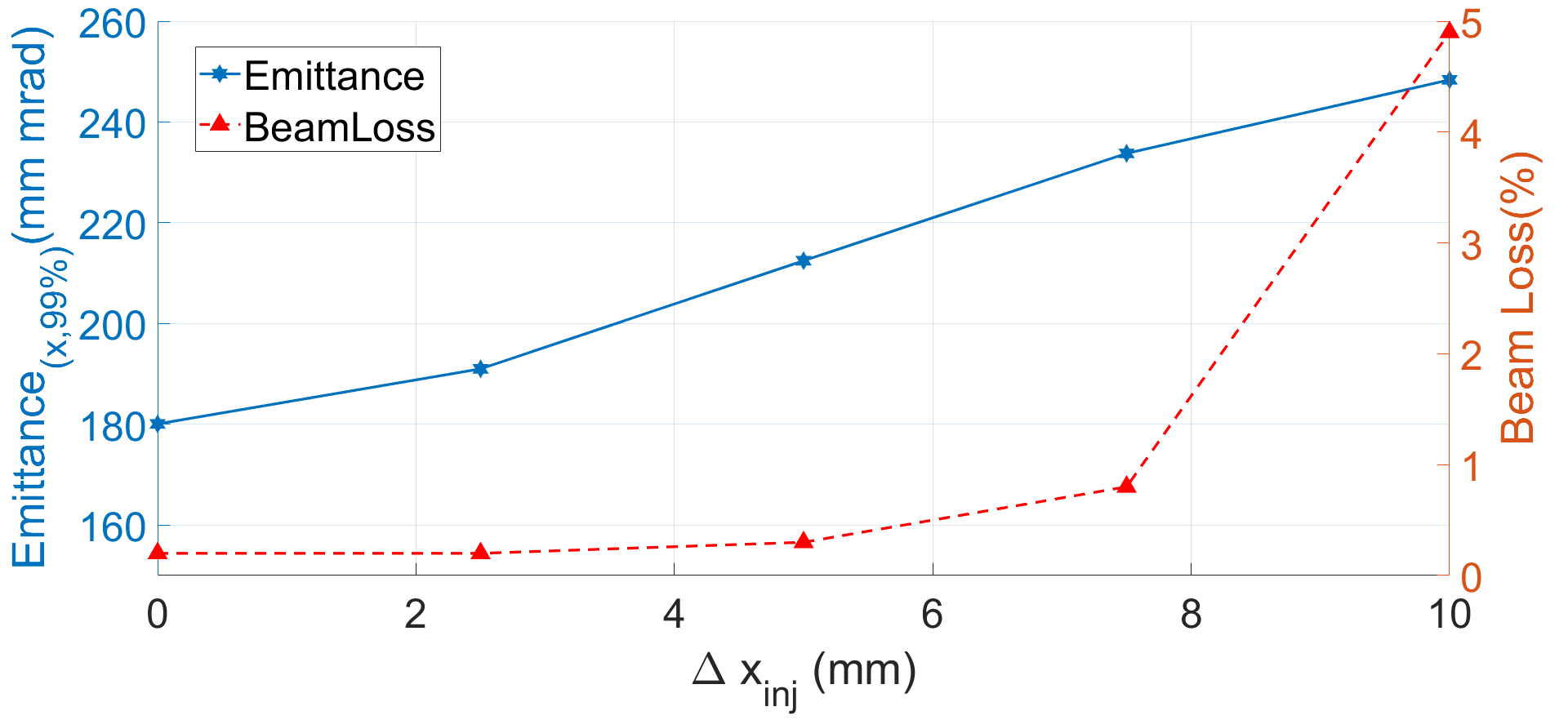}
\caption{\label{figemit} (Color)The emittance growth(blue solid line) and beam loss(red dash line) due to mismatch of injection orbit}
\end{figure}

As shown in Fig. \ref{figemit}, the emittance increases with the increase of $\Delta x$(mismatch) and the beam loss is significant increased when $\Delta x > 5~mm$.

The identification of the relative injection beam orbit is critical to perform the good injection matching. However, the injection orbit belongs to the transport line and the circulating orbit belongs to the ring, and it's hard to measure the relative injection orbit directly. Theoretically, the relative injection beam orbit can be calculated by transfer matrix with turn-by-turn beam position of a single turn injected beam. But for CSNS, the maximum intensity of single turn injected beam is $15~mA$, and it is too low to be measured with sufficient signal-noise ratio by using beam position monitor(BPM). So the beam current should be increased with multi-turn injection to gain in the signal-noise ratio of BPM, while the identification of the phase-space coordinates at the injection point becomes more complicated.

In this paper, two methods based on multi-turn injection and turn-by-turn data of beam position are introduced to identify the injection orbit relative to circulating orbit. The simulation results and application in the beam commissioning of CSNS/RCS show the validity of the methods.

\section{A method to identify the injection orbit based on transfer matrix}
\subsection{Basic principle}
The matrix of betatron oscillation:
\begin{equation}
\label{eq1}
\begin{bmatrix}
\Delta x_i\\
\Delta x_i^{'}
\end{bmatrix}
 = M_i
\begin{bmatrix}
\Delta x_{inj}\\
\Delta_{inj}^{'}
\end{bmatrix},
\end{equation}
where $\Delta x_{inj}$,$\Delta x_{inj}^{'}$ and $\Delta x_i$,$\Delta x_i^{'}$ are the transverse phase space coordinates relative to the circulating orbit at the injection point and the $i$th BPM in the ring respectively, and $M_i$ is the transfer matrix from the injection point to the $i$th BPM。

As the BPM detect the charge center of the beam, for $m$ turns injection beam, the transverse phase-space coordinates at the $j$th BPM after $n$ turns could be derived as:

\begin{equation}
\label{eq2}
\begin{bmatrix}
\Delta x_j\\
\Delta x_{j}^{'}
\end{bmatrix}
= M_{inj-j}\bigl(M^{n-1}+M^{n-2}+\cdots+M^{n-m}\bigr)/m
\begin{bmatrix}
\Delta x_{inj}\\
\Delta x_{inj}^{'}
\end{bmatrix},
\end{equation}
where $\Delta x_j$ and $\Delta x_{j}^{'}$ are the phase-space coordinates of the circulating beam, $\Delta x_{inj}$ and $\Delta x_{inj}^{'}$ are the injection beam coordinates at the injection point, $M_{inj-j}$ is the transfer matrix from injection point to the $j$th BPM, $M$ is the one turn transfer matrix.

In the Eq. (\ref{eq2}), $\Delta x_j$ could be measured directly by using the turn-by-turn BPM, $\Delta x_{j}^{'}$ could be obtained with a pair of BPMs which located in drift space~\citep{ref5}. Here a more general algorithm is introduced to obtain $\Delta x_{inj}$ and $\Delta x_{inj}^{'}$, which is described as:
\begin{eqnarray}
m
\begin{bmatrix}
\Delta x_1\\
\Delta x_1^{'}
\end{bmatrix}
= M_1(n)
\begin{bmatrix}
\Delta x_{inj}\\
\Delta x_{inj}^{'}
\end{bmatrix},
\label{eq3}\\
m
\begin{bmatrix}
\Delta x_2\\
\Delta x_2^{'}
\end{bmatrix}
= M_2(n)
\begin{bmatrix}
\Delta x_{inj}\\
\Delta x_{inj}^{'}
\end{bmatrix},
\label{eq4}
\end{eqnarray}

$=>$
\begin{equation}
m
\begin{bmatrix}
\Delta x_1\\
\Delta x_2
\end{bmatrix}
= \begin{bmatrix}
[M_1(n)]_{11} & [M_1(n)]_{12}\\
[M_2(n)]_{11} & [M_2(n)]_{12}
\end{bmatrix}
\begin{bmatrix}
\Delta x_{inj}\\
\Delta x_{inj}^{'}
\end{bmatrix},
\label{eq5}\\
\end{equation}
where $M_j(n) = M_{inj-j}\bigl(M^{n-1}+M^{n-2}+\cdots+M^{n-m}\bigr)$ is the transfer matrix from injection point to $j$th ($j$=1,2) BPM after n turns circulating.

With the above algorithm, any BPM could be utilized in the measurement of machine study.
\subsection{Numerical simulations}

In the simulation and beam commissioning, the primary errors should be considered are the errors of transfer and BPM errors which related to the resolution. In beam commissioning, the tune deviation due to transfer matrix errors can be controlled to less than 0.005 by making optics measurement and correction with LOCO (Linear Optics from Closed Orbits)~\citep{ref6}. A simulation with transfer matrix errors was done by using Accelerator ToolBox (AT) which is a collection of tools to model particle accelerators in the MATLAB environment~\citep{ref7}, the reconstructed $\Delta x_{inj}$ and $\Delta x_{inj}^{'}$ in Eq.~(\ref{eq5}) are less than $\pm$0.5 $mm$ and $\pm$0.03 $mrad$, when the tune deviation due to transfer matrix errors is less than 0.005. So only the BPM errors are considered in the next simulations.

\begin{figure}
\includegraphics[width=8.5cm]{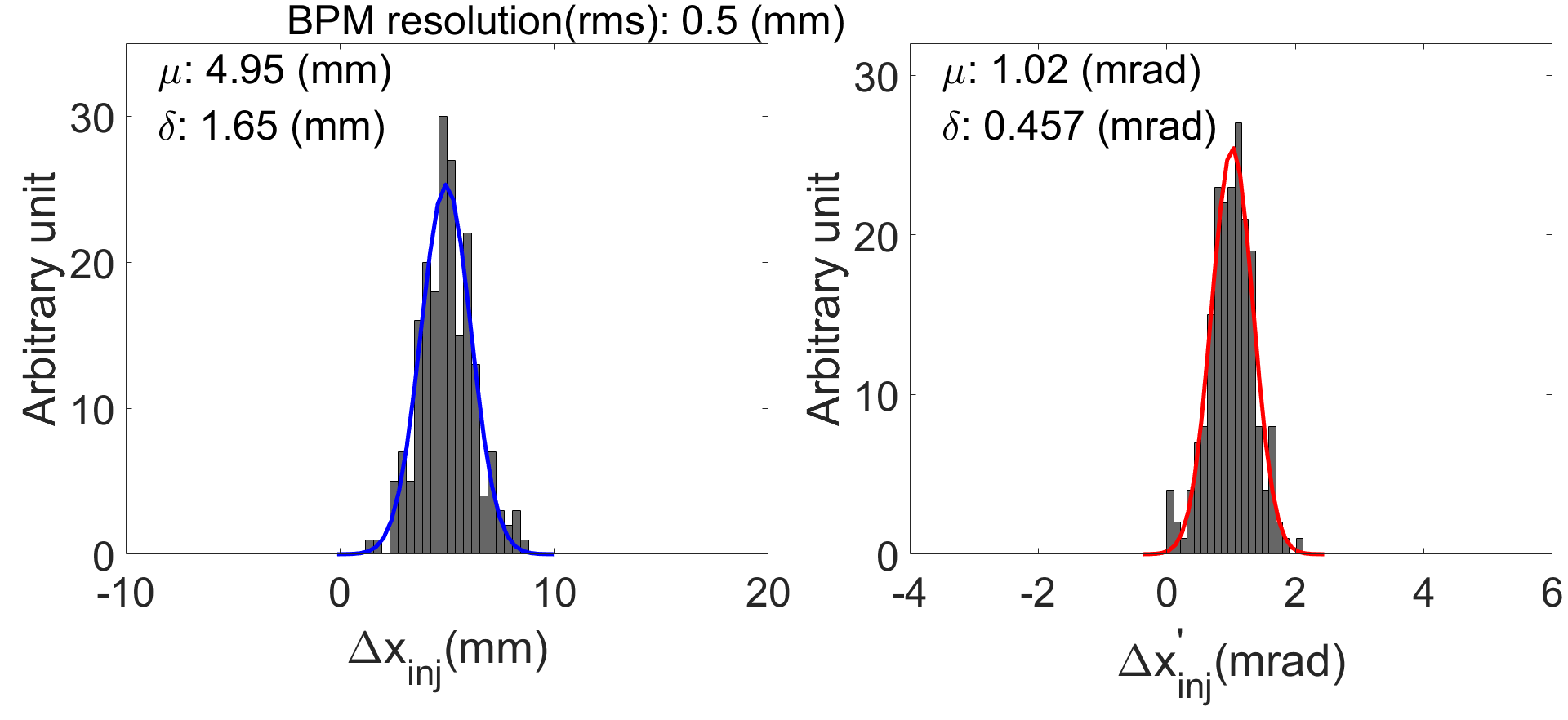}
\includegraphics[width=8.5cm]{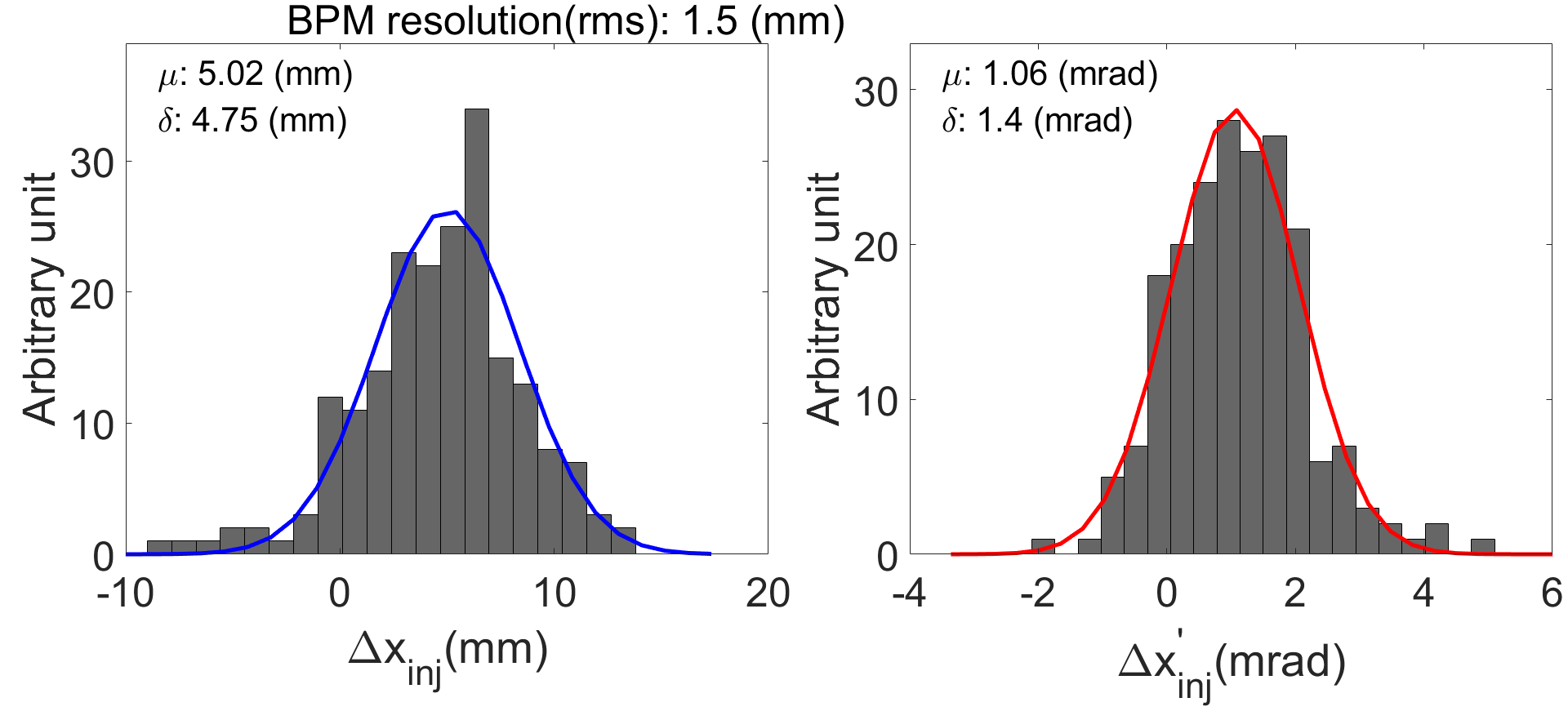}
\includegraphics[width=8.5cm]{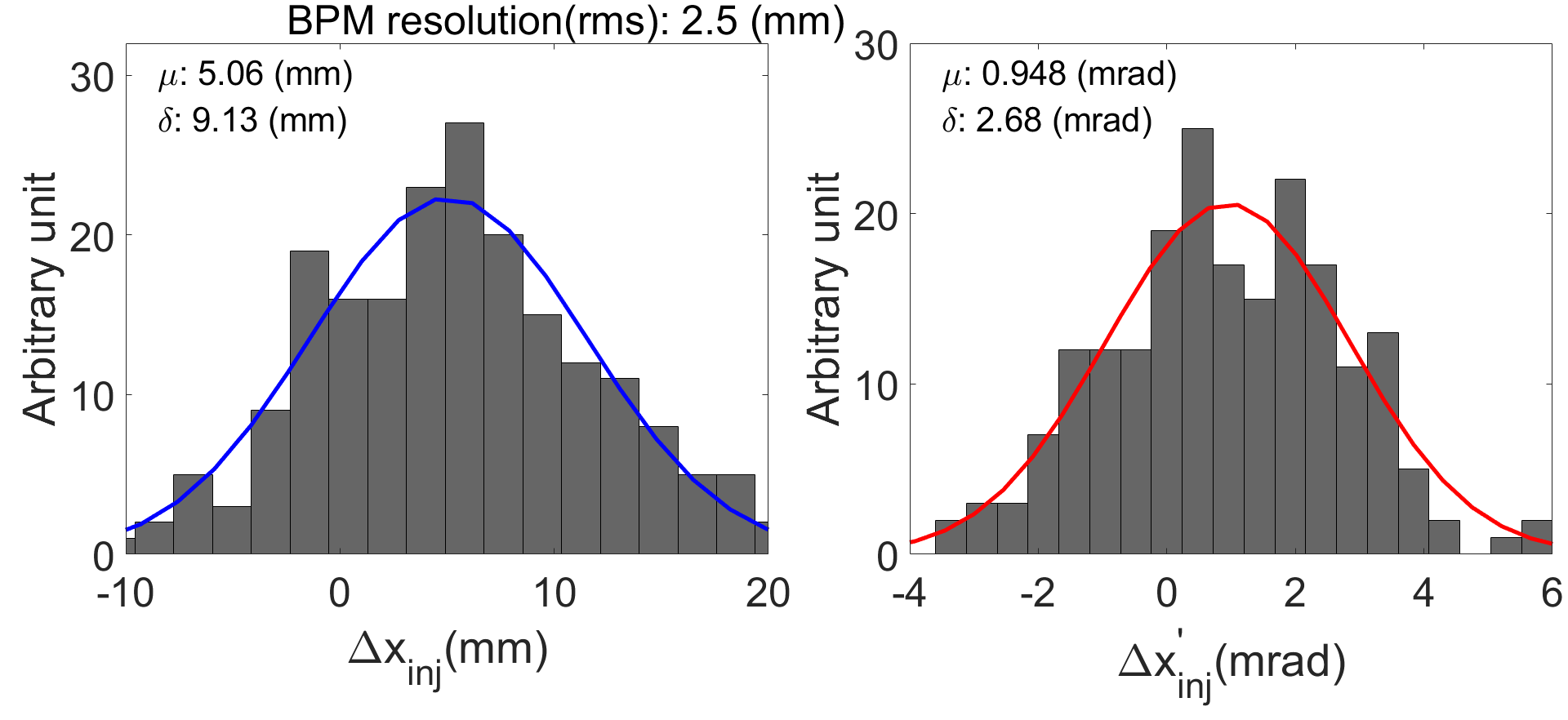}
\caption{\label{fig2}
(Color)The simulation results of reconstructing the phase-space coordinates relative to the circulating orbit at the injection point were obtained by repeating calculations of 200 times with the algorithm described in section 2.1. The left plots are the position of phase-space coordinates and the right plots are the angle. The mean value and the width were obtained by fitting each distribution to a Gaussian function(blue solid line for position and red solid line for angle). The top plot is the result with BPM resolution 0.5 $mm$, the middle and the bottom plots are the results with BPM resolution 1.5 $mm$ and 2.5 $mm$ respectively.}
\end{figure}

In the simulation, the random BPM errors, normally distributed, was generated with different BPM resolution, 0.5 $mm$, 1.5 $mm$ and 2.5 $mm$ respectively. In the simulation, 3 turns beam were injected and the preset horizontal phase-space coordinates at the injection point was (5 $mm$, 1 $mrad$). R4BPM12 and R4BPM02 were marked as the observation point during the tracking. Totally 200 sets of BPM errors were simulated, and the reconstructed phase-space coordinates at the injection point were obtained by using Eq.~(\ref{eq5}). The results were plotted in a histogram as shown in the Fig.~\ref{fig2}, in which the left plots are the position of the phase-space coordinates and the right plots are the angle coordinates. The mean values ($\mu$) and the standard deviation ($\delta$) of the results were obtained by fitting each distribution to Gaussian function. The simulated means of ($\Delta x_{inj}$, $\Delta x_{inj}^{'}$) with different BPM resolution were close to the preset values of (5 $mm$, 1 $mrad$), which reflected the validity of the method. The standard deviations of ($\Delta x_{inj}$, $\Delta x_{inj}^{'}$) were found to be (1.65 $mm$, 0.457 $mrad$), (4.75 $mm$, 1.4 $mrad$), (9.13 $mm$, 2.68 $mrad$) for BPM resolution of 0.5 $mm$, 1.5 $mm$ and 2.5 $mm$ respectively. One can see that the simulation result is very sensitive to the resolution of BPM, and the high BPM resolution is required in this algorithm. For CSNS/RCS, the nominal BPM resolution is about 2 $mm$, and it is difficult to achieve high accuracy result by using this algorithm. To improve the accuracy, an improved algorithm is proposed, as discussed in section 3. 

\section{A method to identify the injection orbit based on fourier transform}
\subsection{Basic principle}
When the injection beam is deviated from the closed orbit of circulating beam at injection point, the betatron oscillation could be detected by BPM. Theoretically, the displacement from the reference orbit after $n$ turns at one specified BPM could be obtained:
\begin{equation}
\label{eq8}
\Delta x(n)=\Delta x_{0}
cos(2\pi n \upsilon_{x})+
(\Delta x_{0}\alpha_{x}
+ \Delta x_{0}^{'}\beta_{x})sin(2\pi n \upsilon_{x}),
\end{equation}
where $\Delta x_{0}$ and $\Delta x_{0}^{'}$ are the initial phase-space coordinates at the specified BPM, $\upsilon_{x}$ is the betatron tune, $n$ is the turn number, $\alpha_{x}$ and $\beta_{x}$ are $twiss$ parameters. Generally, the above oscillation is simple harmonic oscillation with respects to the turn number $n$.
Eq.~(\ref{eq9}) is the fourier transform of beam motion,
\begin{equation}
\label{eq9}
X(\upsilon_{x})= \sum_{n=0}^{N-1}\Delta x(n)e^{-i n \upsilon_{x} T},
\end{equation}
where $T$ is the cycle period of the ring. Assume that
\begin{eqnarray}
C(\upsilon_{x}) = \sum_{n=0}^{N-1}
\bigl\{
cos(2\pi n \upsilon_{x})+
\alpha_{x}sin(2\pi n \upsilon_{x})
\bigr\}
e^{-i n \upsilon_{x} T},\\
S(\upsilon_{x}) = \sum_{n=0}^{N-1}
\beta_{x}sin(2\pi n \upsilon_{x})
e^{-i n \upsilon_{x} T},
\end{eqnarray}
then Eq.~(\ref{eq9}) could be expressed as:
\begin{equation}
\label{eq12}
X(\upsilon_{x})= C(\upsilon_{x})\Delta x_0+S(\upsilon_{x})\Delta x_0^{'}.
\end{equation}

In the Eq.~(\ref{eq12}) the real part $Re[X(\upsilon_{x})]$ and imaginary part $Im[X(\upsilon_{x})]$ depend on the corresponding parts of $C(\upsilon_{x})$ and $S(\upsilon_{x})$, which can be described as:
\begin{equation}
\label{eq13}
\begin{bmatrix}
Re[X(\upsilon_{x})]\\
Im[X(\upsilon_{x})]
\end{bmatrix}
=R(\upsilon_{x})
\begin{bmatrix}
\Delta x_0\\
\Delta x_0^{'}
\end{bmatrix},
\end{equation}
where
\begin{equation}
R(\upsilon_{x})
=
\begin{bmatrix}
Re[C(\upsilon_{x})] & Re[S(\upsilon_{x})]\\
Im[C(\upsilon_{x})] & Im[S(\upsilon_{x})]
\end{bmatrix},
\end{equation}

Based on the features of fourier transform, $Re[X(\upsilon_{x})]$ is actually the amplitude of the cosine component and $Im[X(\upsilon_{x})]$ is the amplitude of the sine component, the response matrix could be described as follow:
\begin{equation}
R(\upsilon_{x})
=
\begin{bmatrix}
1 & 0\\
\alpha_{x} & \beta_{x}
\end{bmatrix}.
\end{equation}

According to Eq.~(\ref{eq2}), the transfer matrix from injection point to one specified BPM after $n$ turn with $m$ turn injection beam could be obtained as:
\begin{equation}
\label{eq14}
\begin{bmatrix}
\Delta x_0\\
\Delta x_0^{'}
\end{bmatrix}
= M_{m/n}
\begin{bmatrix}
\Delta x_{inj}\\
\Delta x_{inj}^{'}
\end{bmatrix},
\end{equation}
where
\begin{equation}
M_{m/n}=M_{inj-j}\bigl(M^{n-1}+M^{n-2}+\cdots+M^{n-m}\bigr)/m.
\end{equation}

The real and imaginary parts of $X(\upsilon_{x})$ could be described by matrix $A=R(\upsilon_{x})M_{m/n}$ and the initial phase-space coordinates at injection point:

\begin{equation}
\label{eq15}
\begin{bmatrix}
Re[X(\upsilon_{x})]\\
Im[X(\upsilon_{x})]
\end{bmatrix}
=
A
\begin{bmatrix}
\Delta x_{inj}\\
\Delta x_{inj}^{'}
\end{bmatrix},
\end{equation}
$\upsilon_{x}$ could be obtained from $turn-by-turn$ data with fourier analysis, and the matrix $A$ could be obtained by beam based measurement.

Then the phase-space coordinates of injected beam relative to the circulating beam at the injection point could be obtained with the turn-by-turn data of a single BPM.

\subsection{Numerical simulations}

\begin{figure}
\includegraphics[width=8.5cm,height=3.5cm]{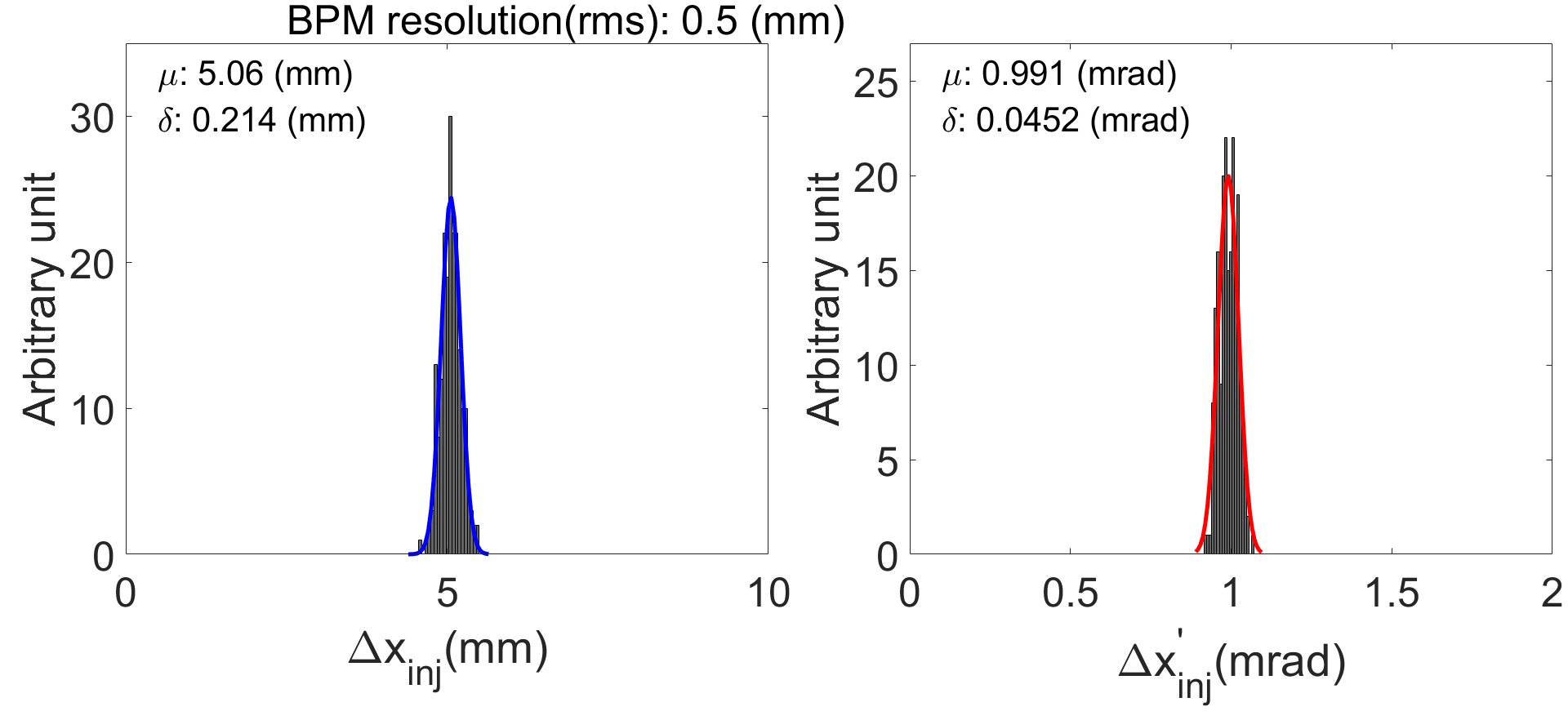}
\includegraphics[width=8.5cm,height=3.5cm]{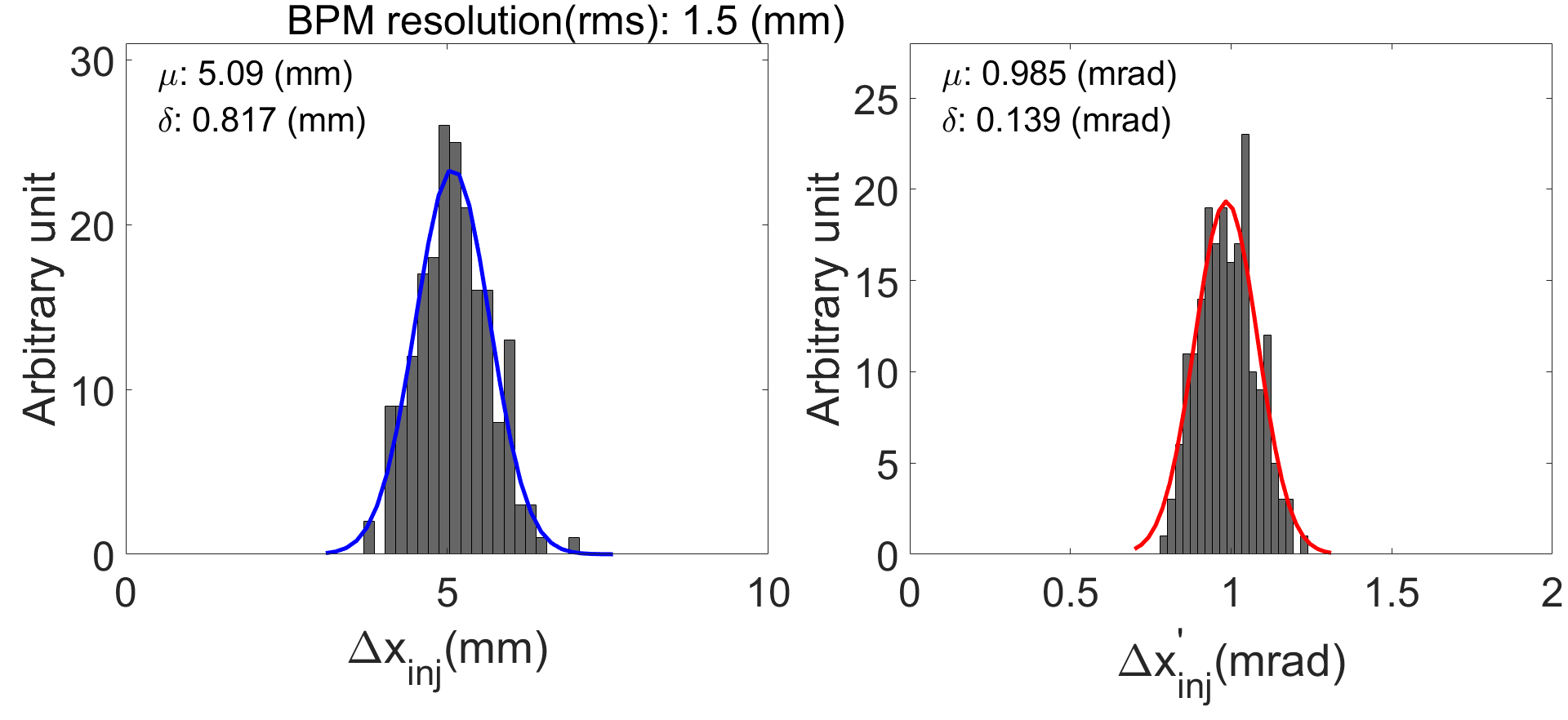}
\includegraphics[width=8.5cm,height=3.5cm]{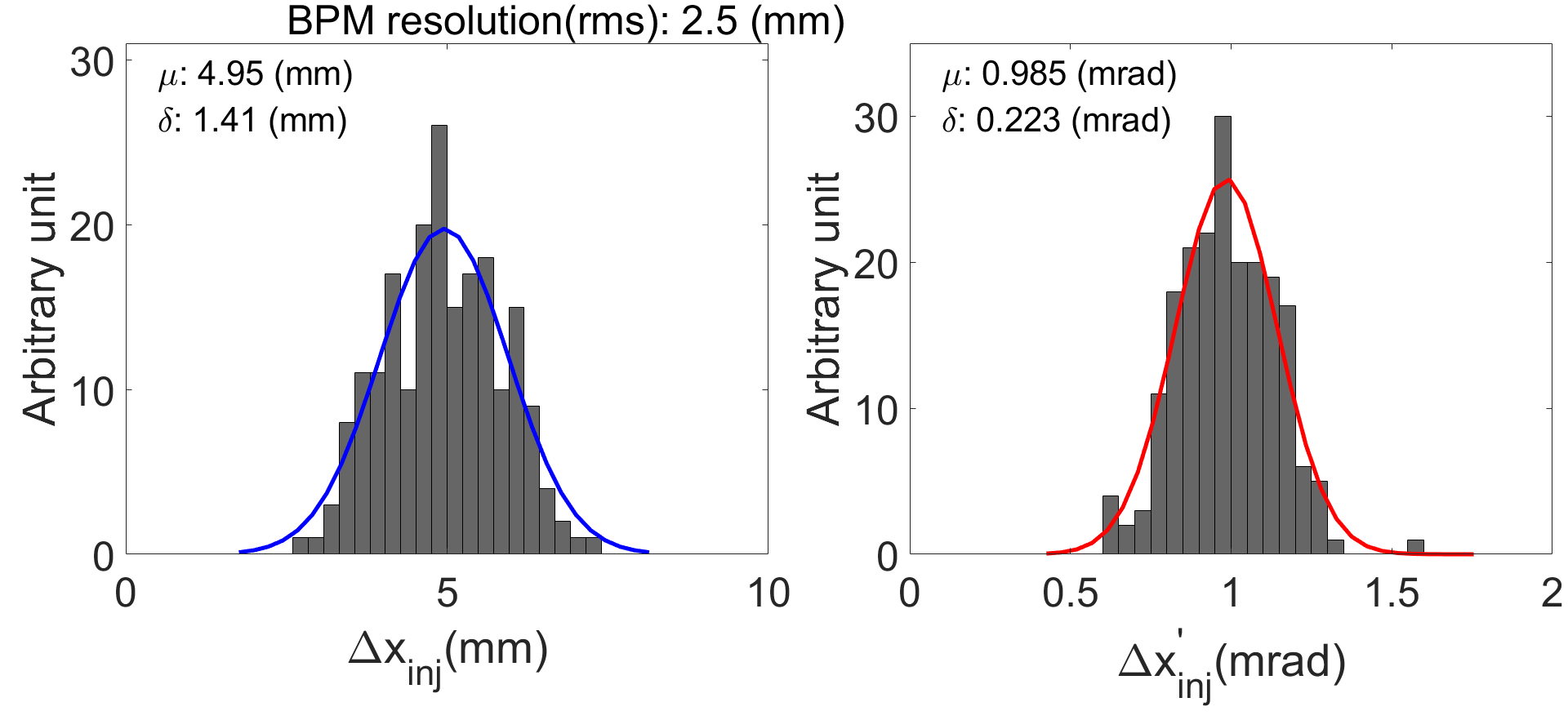}
\caption{\label{fig5} (Color) Similar to Fig.~\ref{fig2} but the tracking turns was increased to 20, and R4BPM12 was marked as the observation point during tracking.}
\end{figure}

The simulation, similar as what done in section 2.2, was performed. Most of the simulation conditions were consistent with that of the simulation in section 2.2, except for that the tracking turn was increased to 20.
The simulation results are shown in Fig.~\ref{fig5}. For the different preset errors, the mean values of ($\Delta x_{inj}$, $\Delta x_{inj}^{'}$) are consistent and close to the preset values (5 $mm$, 1 $mrad$), which indicated the validity of the algorithm. The standard deviations of ($\Delta x_{inj}$, $\Delta x_{inj}^{'}$) were (0.214 $mm$, 0.0452 $mrad$), (0.817 $mm$, 0.139 $mrad$), (1.41 $mm$, 0.223 $mrad$) for BPM resolution of 0.5 $mm$, 1.5 $mm$, 2.5 $mm$ respectively, and which is quite small even with the large error of BPM resolution of 2.5 $mm$. The simulation shows the good accuracy of the method.

So far, all the analysis and simulations are based on the linear condition. As mentioned in section 2.1, the BPM detect the charge center of the beam, so under the linear condition the beam position of the multi-turn injected beam is equal to the average of the beam positions of each single pulse. However, in reality, a few factors will break the linear condition, in which the space charge is one of the main non-linear factors. A simulation was performed to evaluate the effect of space charge on the beam position with multi-turn injection. 

\begin{figure}
\includegraphics[width=8cm]{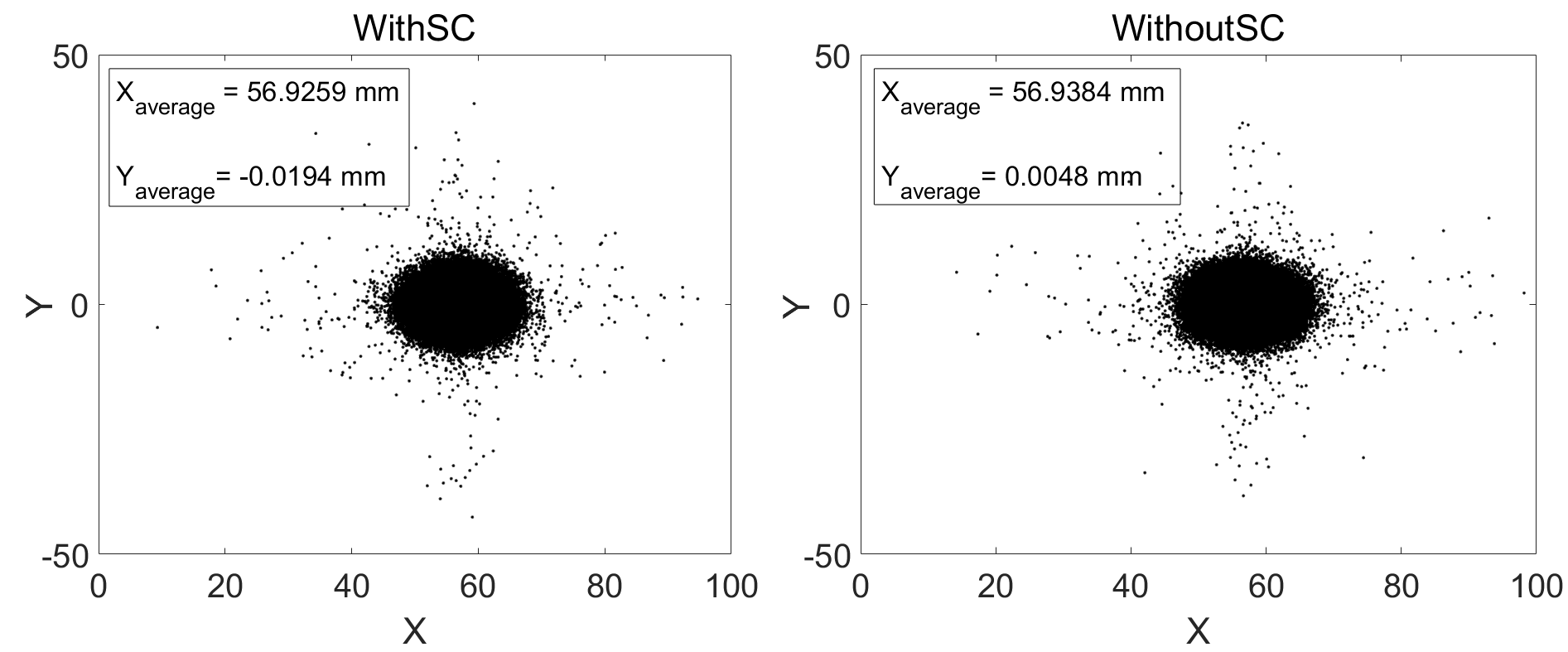}
\caption{\label{figsc}The comparison of the tracking results with and without space charge. The left plot is the tracking result with space charge, and the right plot is the tracking result without space charge.}
\end{figure}

The process of the injection with space charge and without space charge were simulated by ORBIT respectively. The turn number of the injected beam was 3 and each turn of injected beam contained 40000 macro particles which corresponded to the peak current of 10 $mA$, and the first 20 turns were considered in simulation. The simulation results are shown in the Fig. \ref{figsc}, in which the left plot shows the tracking result in real space with space charge while the right plot shows the tracking result without space charge. It can be seen that for the 3 turns injected beam, and with peak current of 10 $mA$ , after 20 turns of tracking, the effects of the space charge on the beam distribution, as well as the beam position, is very small, and can be ignored.

\section{Machine study}

The method was tested and applied in the beam commissioning of CSNS/RCS. The beam commissioning was still in the beginning state, and the closed orbit wasn't well corrected, so the machine study was conducted for only the second method.

The machine study was performed in the DC mode of RCS, in which the beam was injected and extracted in 80 $MeV$ without acceleration. The harmonic number of RCS is 2, but in the machine study, the chopped beam was only injected into one of two buckets, and totally 3 pulses from the linac were injected into the RCS with the peak current of 12 $mA$. Two RF cavities were utilized to provide 24 $kV$ voltage to the beam, and the synchrotron period is about 0.36 $ms$. After the injection , the first 18 turns BPM data were taken for utilization in calculation. The circulating period is about 1.96 $\mu s$, and the 18 turns circulating take about 0.035 $ms$, and is far less than the period of synchrotron oscillation, so the effect of synchrotron oscillation on the transverse position of the beam is negligible. The BPMs which in the dispersion-free area were selected to avoid the effect of the dispersion. The matrix A which was utilized in the calculation and described in section 3.1 was obtained from a beam calibrated model based on AT, the measured tune in horizontal direction was 4.822 and the tune calculated by the calibrated model was about 4.824, according to the simulation result in section 2.2, the errors of transfer matrix could be ignored.

\begin{figure}
\includegraphics[width=8.5cm,height=3cm]{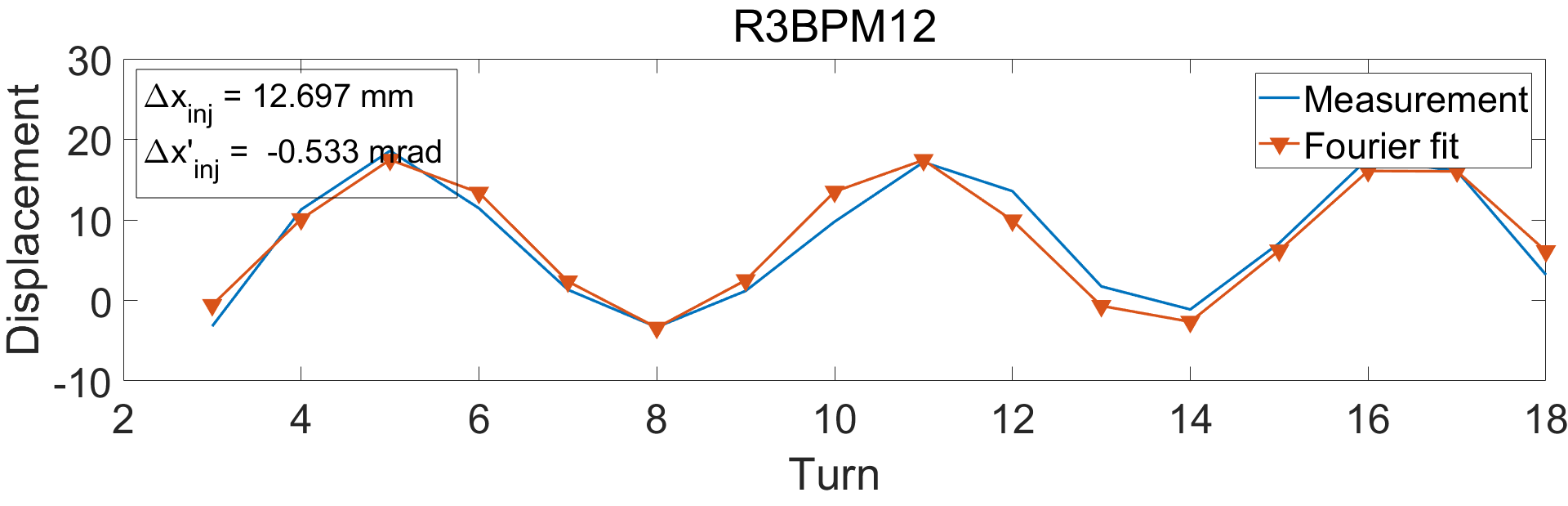}
\includegraphics[width=8.5cm,height=3cm]{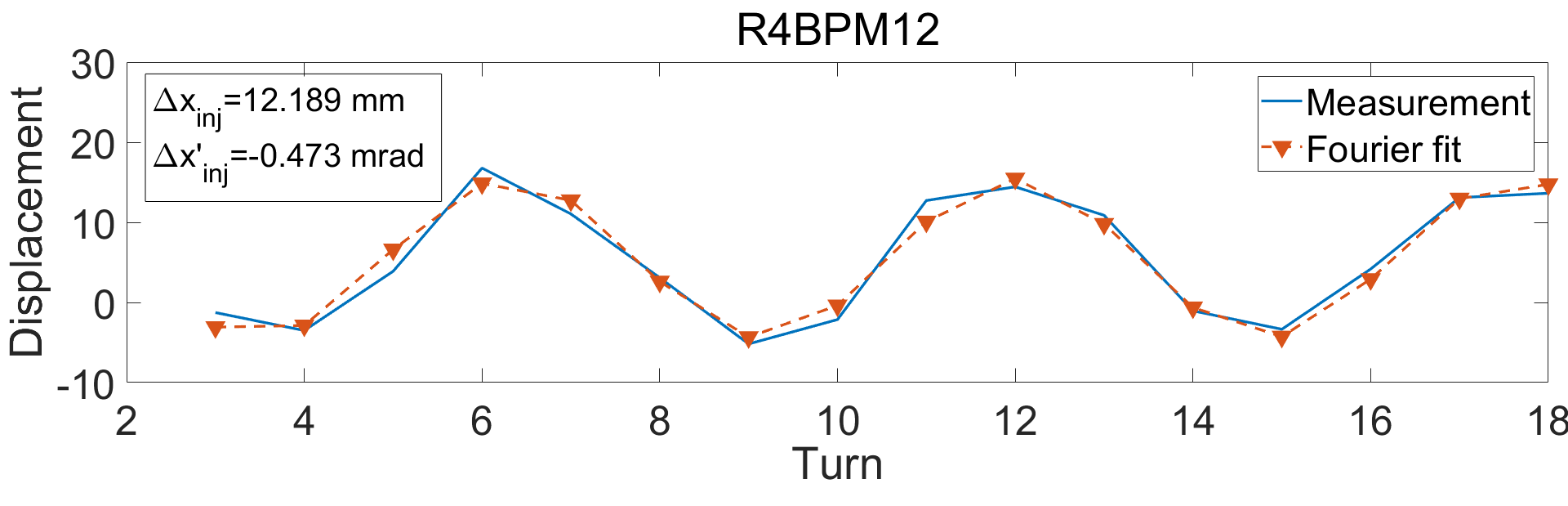}
\caption{\label{figmeasure} (Color)The measurement results of betatron oscillation at two different BPMs (blue solid lines)and their fourier fit(red dash lines with triangles).}
\end{figure}

Two BPMs were selected as the observation points. Fig.~\ref{figmeasure} shows the measurement results, and in which the betatron oscillation can be clearly observed.

The phase-space coordinates of injected beam at the injection point were reconstructed with turn-by-turn data from different BPMs, as listed in Table~\ref{tabres}. Considering the resolution of BPM, the consistency of the results obtained from two different BPMs is satisfactory.

\begin{table}
\caption{ \label{tabres} The results of the reconstructed phase space coordinates at the injection point which based on the measured data.}
\begin{ruledtabular}
\begin{tabular*}{75mm}{c@{\extracolsep{\fill}}c}
\textbf{BPM} & \textbf{Injection $(\Delta x_{inj},\Delta x_{inj}^{'})$} $(mm, mrad)$\\
\colrule
R3BPM12 & (12.697, -0.533) \\
R4BPM12 & (12.189, -0.473) \\
\end{tabular*}
\end{ruledtabular}
\end{table}

\begin{figure}
\includegraphics[width=8.5cm,height=3cm]{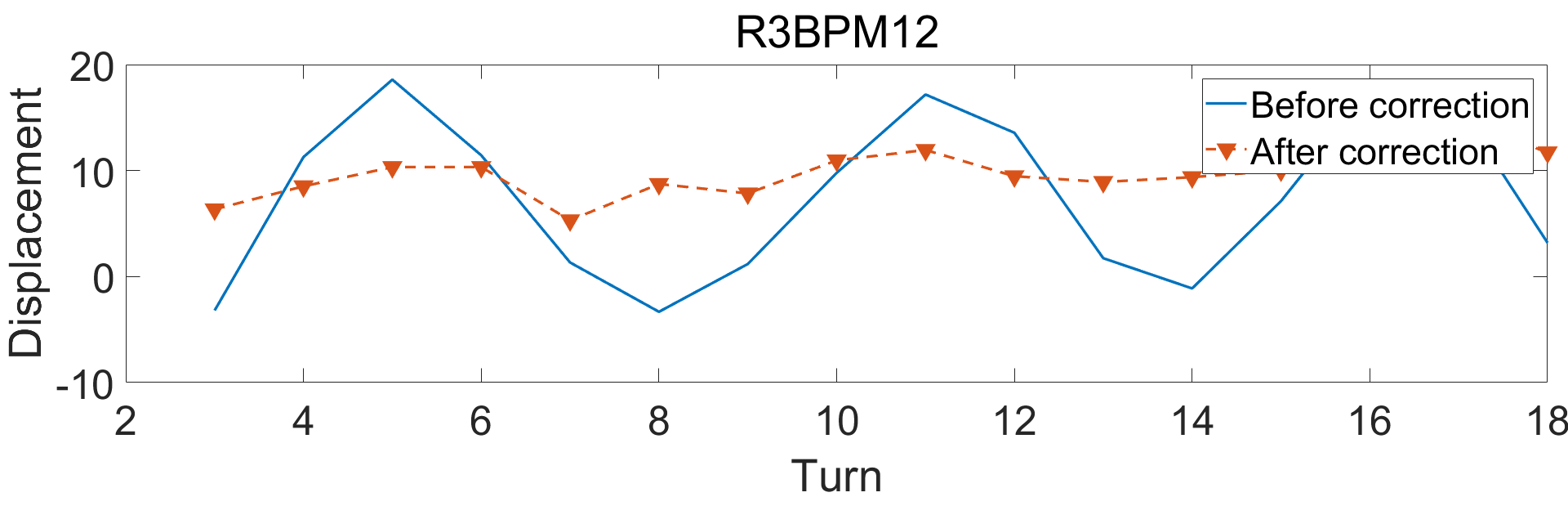}
\includegraphics[width=8.5cm,height=3cm]{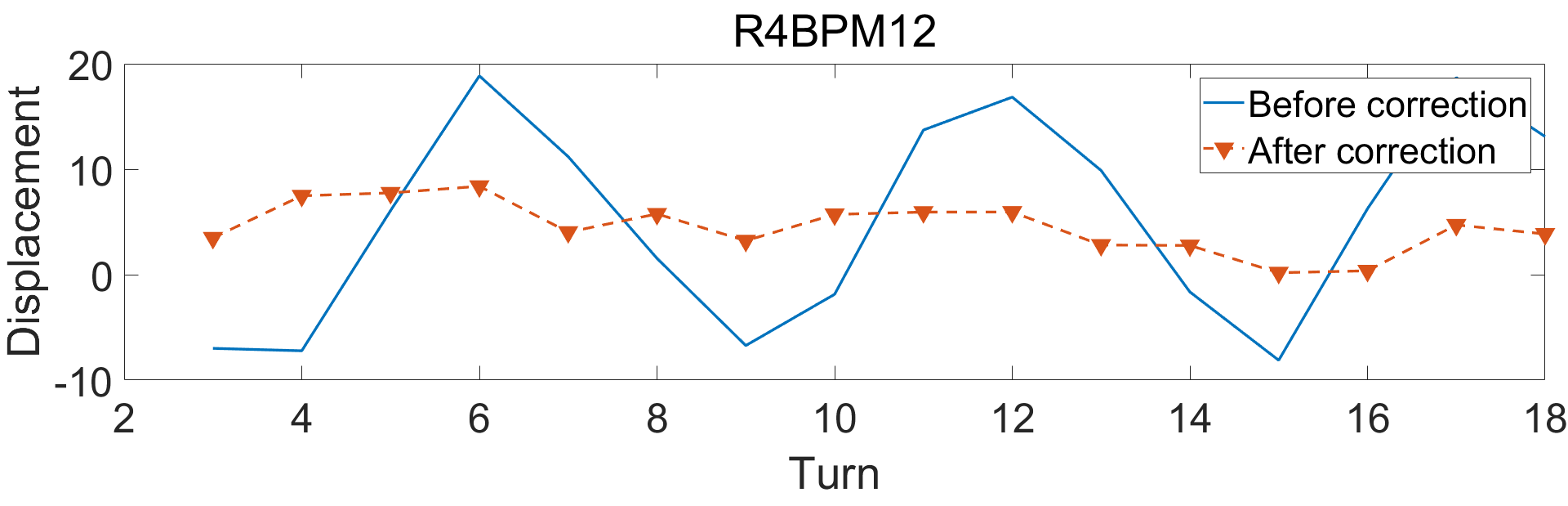}
\caption{\label{figcorr} (Color)The detected betatron oscillation at two different BPMs before (blue solid line) and after (red dash line with triangles) correction of deviated coordinates of injection beam.}
\end{figure}

The relative position and angle of the injected beam at injection point can be corrected by the bump magnets in RCS and dipole correctors in the beam transport line. When the coordinates deviation of injected beam were corrected, with 3 turn injection beam, the oscillation of beam position in the BPMs can be depressed. Fig.~\ref{figcorr} shows the detected betatron oscillation before and after correction.

\begin{figure}
\includegraphics[width=8.5cm]{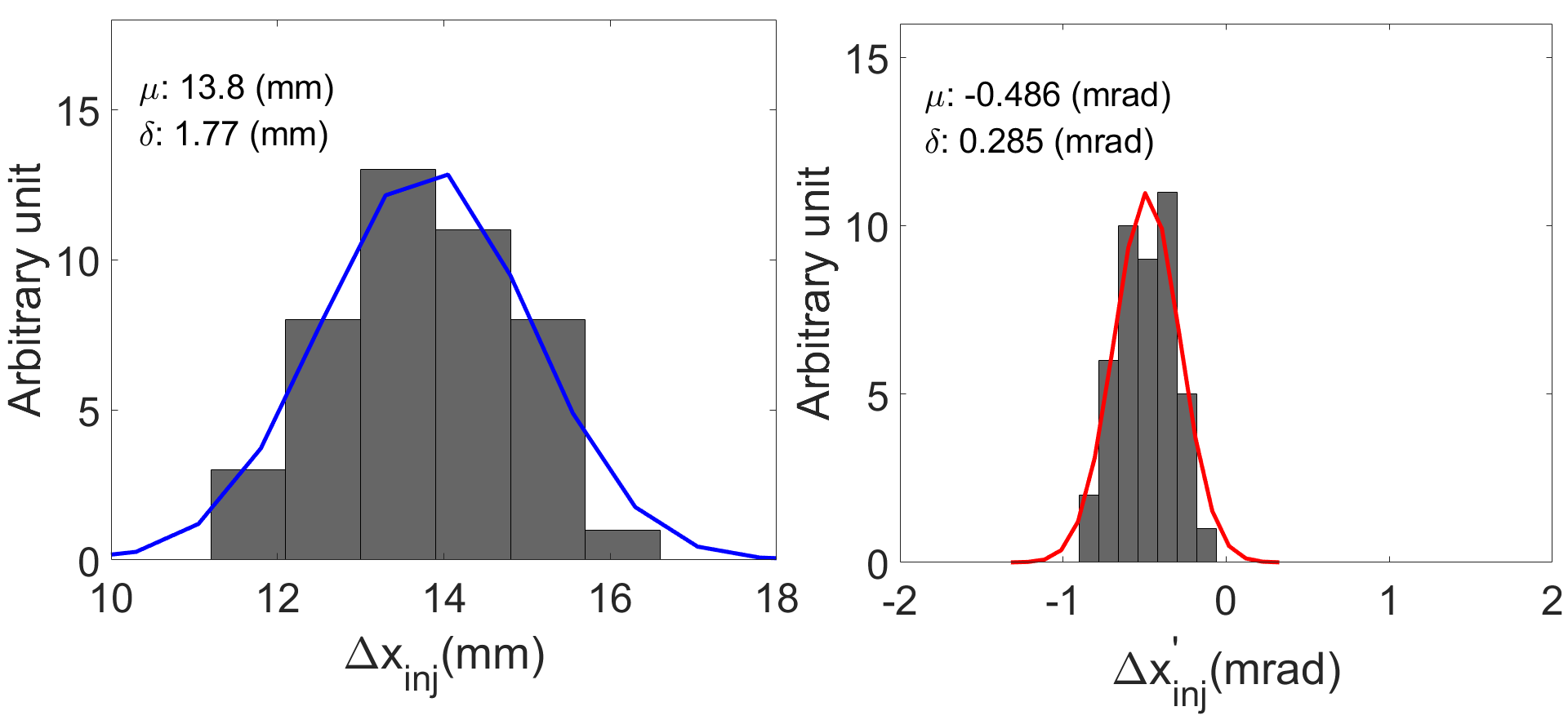}
\caption{\label{figstat} (Color) The overall errors of the results which were obtained using the second method by repeating measurements 11 times.}
\end{figure}

Since the method is based on the turn-by-turn data of the BPMs and transfer matrix, the accuracy of the results depends on the resolution of the BPMs and errors of transfer matrix. However, in practice, the stability of the injection beam and the fluctuation of the bump magnets are also considerable factors. In the machine study, it is difficult to evaluate the effect of these factors individually. However, the overall influence of the fluctuation on the measurements could be extracted by repeatedly do the measurement. 
Fig.~\ref{figstat} shows the statistics results of 11 times of beam best, and the turn-by-turn data were obtained from 4 BPMs . The left plot is the position of phase-space coordinates and the right plot is the angle. The mean values of ($\Delta x_{inj}$, $\Delta x_{inj}^{'}$) were found to be 13.8 $mm$ and -0.486 $mrad$,  respectively. The standard deviations of ($\Delta x_{inj}$, $\Delta x_{inj}^{'}$), here are 1.77 $mm$ and 0.285 $mrad$, could be used to estimate the effects of the overall errors.

\section{Discussion}
These two methods were performed validly by simulation, and the second method was applied in the beam commissioning successfully. The phase-space coordinates relative to the circulating orbit at the injection point can be reconstructed effectively. However, there is obvious difference between the accuracy of the two methods. The first method is very sensitive to the BPM errors, and the accuracy of the result depends on both the BPM resolution and offset. The reason is that the errors  are involved in calculation directly, and there isn't sub-process to depress the effects of the BPM errors. The second method is more robust, it's relatively not sensitive to the errors, and the standard deviation of the results is quite small with the relatively low BPM resolution. The key point is that the fourier fitting is utilized to calculate the real and imaginary parts of oscillation data, and the effect of BPM errors are depressed during the fitting.

Compare with the method which based on the single pass mode and BPM pairs, the present two methods based on multi-turn injection are more general. Any BPM of RCS could be marked as observation point in the calculation, and the accumulation of beam intensity could enhance the signal noise ratio of BPM data. 

\section{Conclusion}
In the painting injection, the mismatch between the injection orbit and the circulating orbit will result in emittance growth and non-uniformity of beam distribution, which may lead to beam loss. Due to the special location and limitation of detector,  the identification of the phase-space coordinates of the injection beam at injection point is rather difficult. Two effective methods were introduced to obtain the phase-space coordinates at injection point base on the multi-turn injection and turn-by-turn data of beam position, and both methods are checked by simulation study. Due to the limit of accuracy and resolution of BPM, only the second method was applied in the beam commissioning of CSNS/RCS successfully, and the beam test results were satisfactory. The detected betatron oscillation was depressed obviously after correction, which means the injection orbit and circulating orbit were well matched.

The beam commissioning of CSNS/RCS is still in the beginning state, the further study will be conducted in the future to enhance the accuracy of the present methods for the application in the painting control. 

\section{Acknowledgements}
The authors want to thank T.G.Xu, S.Y.Xu, J.L.Sun, P.Zhu and other CSNS colleagues for the discussion and consultations.

\clearpage
\end{CJK}

\begin{thebibliography}{90}


\bibitem{ref1}S.Wang et al., Chin Phys C, 33 (2009) 1-3.
\bibitem{ref2}J.Wei et al., Chin Phys C, 33(2009)1033-1042.
\bibitem{ref3-1} J.Y.Tang et al., Chin Phys C, 30(2006) 1184-1189.
\bibitem{ref3}M.Y.Huang et al., Chin Phys C, 37(2013) 067001.
\bibitem{ref4}J.Gabambos et al., ORBIT User's Manual,SNS/ORNL/AP Technical Note 011.1999.
\bibitem{ref5}P.K.Saha,et al.,Physical Review Special Topics: Accelerators and Beams 12(2009)040403
\bibitem{ref6}J.Safranek, G.Portmann, A.Terebilo, "MATLAB-BASED LOCO", Proceedings of EPAC 2002, Paris, France.
\bibitem{ref7}A.Terebilo, "Accelerator Toolbox for Matlab", SLAC-PUB8732, May 2001, http://www.slac.stanford.edu/grp/ssrl/spear/at/

\end{thebibliography}
\end{document}